\documentclass[aps,prl,twocolumn,superscriptaddress,longbibliography]{revtex4-1}
\usepackage{graphicx}
\usepackage{amsmath}
\usepackage{mathtools}
\usepackage[normalem]{ulem}
\usepackage{enumitem}
\usepackage[dvipsnames]{xcolor}
\usepackage[toc,page]{appendix}
\usepackage{listings}

\usepackage[utf8]{inputenc}
\usepackage{comment}
\usepackage{physics}

\usepackage{hyperref}

\begin{document}

\title{Dynamic correlations in Calogero-Sutherland model}

\author{G. Lleopart}
\affiliation{Departament de Ci\'encia de Materials i Qu\'imica F\'isica and Institut de Qu\'imica Te\'orica i Computacional (IQTC), Universitat de Barcelona, c/ Mart\'i i Franqu\'es 1-11, 08028 Barcelona, Spain}
\author{D. M. Gangardt}
\affiliation{School of Physics and Astronomy, University of Birmingham, Edgbaston, Birmingham B15 2TT, United Kingdom}
\author{M. Pustilnik}
\affiliation{School of Physics, Georgia Institute of Technology, Atlanta, GA 30333, USA}
\author{Grigory E. Astrakharchik}

\affiliation{Departament de F\'{\i}sica, Campus Nord  B4-B5, Universitat Polit\`ecnica de Catalunya, E-08034 Barcelona, Spain}
\affiliation{Departament de F{\'i}sica Qu{\`a}ntica i Astrof{\'i}sica, Facultat de F{\'i}sica, Universitat de Barcelona, E-08028 Barcelona, Spain}
\affiliation{Institut de Ci{\`e}ncies del Cosmos, Universitat de Barcelona, ICCUB, Mart{\'i} i Franqu{\`e}s 1, E-08028 Barcelona, Spain}
\date{\today}
\begin{abstract}
The Calogero-Sutherland model represents a paradigmatic example of an integrable quantum system with applications ranging from cold atoms to random matrix theory.
Combining sum rules with the Monte Carlo technique, we introduce a stochastic method that allows one to compute the dynamic structure factor and obtain an exact description of excitations beyond the conventional Luttinger liquid regime.
We explore a broad range of interaction regimes, including weak interactions, where a Bogoliubov-type spectrum emerges, the Tonks-Girardeau regime, where excitations resemble those of an ideal Fermi gas, and strong interactions, where umklapp scattering leads to a Brillouin zone structure, typical of a crystal.
Additionally, we discuss the connection between the hydrodynamic description of one-dimensional quantum gases, liquids, and solids with the Calogero-Sutherland wave function.
The model’s universality extends beyond atoms in waveguides, with implications for disordered systems and random matrix theory.
\end{abstract}
\maketitle

{\it Introduction.}---The quest for exact solutions in many-body quantum systems presents a significant challenge due to the difficulty in solving the Schrödinger equation for $N$ particles. 
Various numerical approaches (such as the diffusion\cite{BoronatCasulleras94} and the Path Integral\cite{CeperleyRMP1995} Monte Carlo methods) allow algorithmically exact evaluation of ground-state properties. 
However, currently, there are no generic methods capable of obtaining the frequency response of many-body systems. 
A common strategy is to calculate the imaginary time correlation functions and apply the Inverse Laplace Transform\cite{Mishchenko12,Goulko17,Bertaina2017}, but unfortunately such transform is a mathematically ill-posed problem as any noise in the imaginary-time data exponentially increases during the evaluation of the real frequency response, preventing its unique determination. 
However, knowledge of the frequency response is crucial for understanding the crossover between the conventional\cite{Haldane81} and the so-called non-linear\cite{PhysRevLett.96.196405,Imambekov2012} Luttinger liquid behavior.
In this work, we aim to address this open problem.

While well-known analytical solutions exist for ideal Bose and Fermi gases\cite{PitaevskiiStringariBook}, exact solutions for quantum many-body systems with finite interactions are far less common.
One-dimensional geometry is exceptional in this regard, as it offers several exactly solvable models\cite{sutherland2004beautiful}.
For example, the wave function of impenetrable bosons of zero (Tonks-Girardeau gas) or finite (hard rods) diameter\cite{Girardeau60} can be mapped to that of an ideal Fermi gas. 
Quantum gases interacting via the contact potential can also be solved using the Bethe {\it ansatz}\cite{LiebLiniger63} method, as particles move freely except for collisions at the contact points.
However, the exact solutions for extended interaction potentials are far less common.
In this context, the Calogero-Sutherland model (CSM) stands out as a remarkable exception\cite{sutherland2004beautiful}. 
It describes one-dimensional particles (bosons or fermions) interacting via $\propto 1/r^2$ interaction potential, its eigenfunctions, in particular the ground state, can be found exactly\cite{Calogero69,Sutherland71}. 
The key feature of this model is that the potential energy $\propto 1/r^2$ scales in the same way as the kinetic energy operator $\partial^2 / \partial r^2$. 
As a result, CSM has a scaling property, that is, by changing the density, it is not possible to change the physical state, which instead is governed by a single dimensionless parameter, the interaction strength $\lambda$. 
As we will argue below, the CSM exhibits universal properties as $\lambda$ plays the role of the Luttinger parameter\cite{Haldane81, Cazalilla04} $K=1/\lambda$ and, in the hydrodynamic regime, the ground state wave function of {\em any} single-component gapless system\cite{ReattoChester67} in one dimension takes the CSM asymptotic form.  
The CSM has also been found to be relevant for spin chains, finding similarities with other theoretical approaches such as Haldane-Shastry theory\cite{CiracSierra10,NielsenCiracSierra11}.

\begin{figure}[!h]
\centering
\includegraphics[width=\columnwidth]{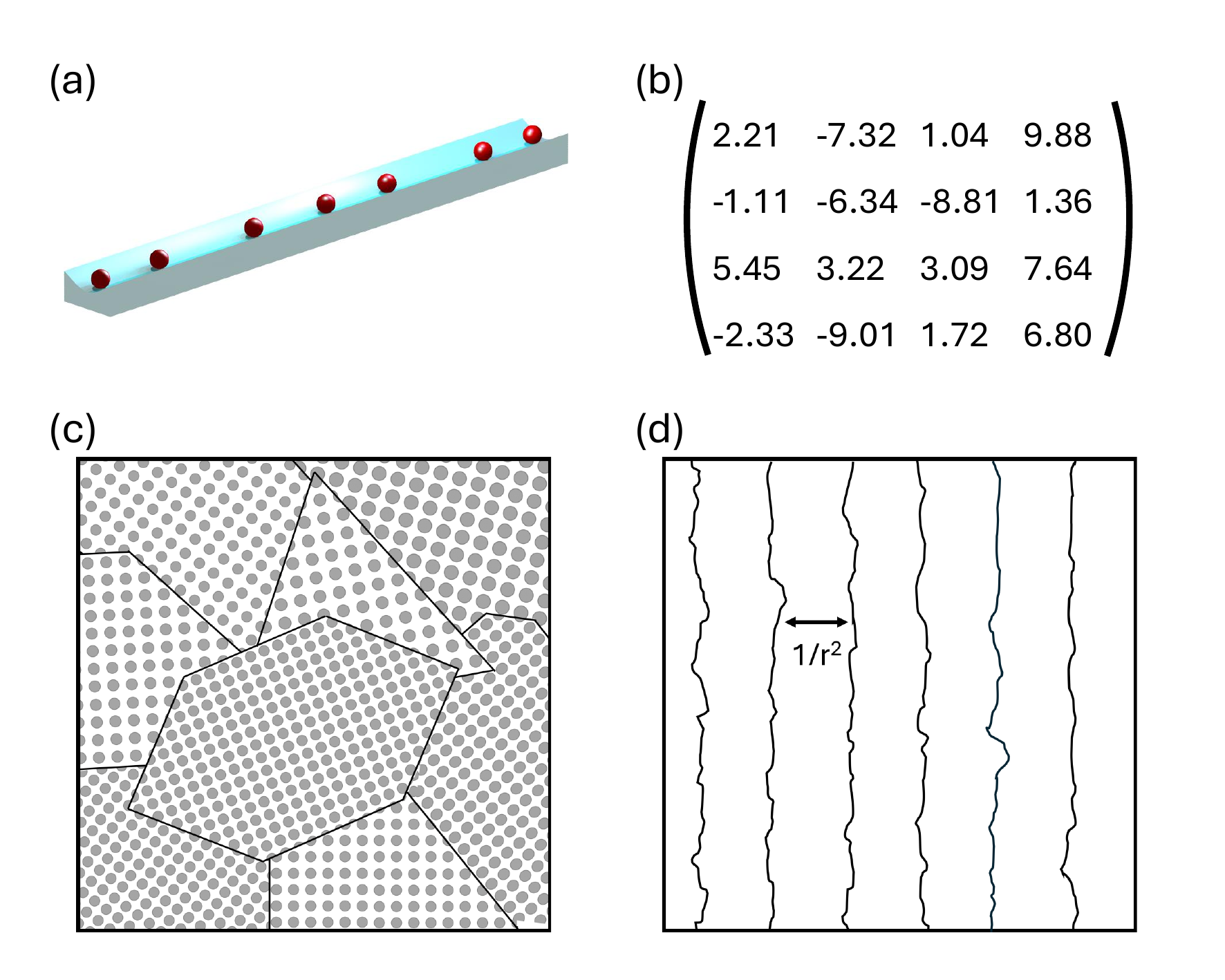}
\caption{Sketch showing contexts in which the Calogero–Sutherland model arises: (a) compressible one-dimensional gases, liquids, and solids at ultralow temperatures; (b) statistics of random matrices; (c) metal grains; and (d) due to effectively $1/r^2$ interactions at small miscut angles, steps on a crystal surface can be mapped onto the worldlines of particles in a one-dimensional quantum system, where the directions perpendicular and parallel to the steps correspond to the spatial coordinate $x$ and imaginary time $\tau$, respectively.
}
\label{fig1}
\end{figure}

Beyond intrinsic significance in quantum mechanics, the Calogero Sutherland model holds substantial interdisciplinary importance, see Fig.~\ref{fig1}. It has explicit connections with Random Matrix theory, as the probability distribution of the level spacing follows the Wigner-Dyson statistics\cite{Wigner51,Dyson62} and it has the same form as the probability distribution function of trapped particles interacting via $1/r^2$ potential. The dimensionless parameter is known as Dyson index $\beta$ in that context, $\beta=2\lambda$, characterizing the orthogonal $\beta=1$, unitary $\beta=2$, and symplectic $\beta=4$ ensembles\cite{SimonsLeeAltshuler93}.

The world lines of CSM in imaginary time can be mapped to the steps on the crystal surface, see Fig.~\ref{fig1}d, which have an effective $1/r^2$ interaction between them\cite{Lassig96} in the limit of small miscut angles. In that mapping, the dimensionless parameter $\lambda$ corresponds to the ratio of the coupling constant and the square of the temperature.

{\it Relation to Luttinger Liquids.}---A powerful effective method known as the Luttinger Liquid theory\cite{Haldane81} exists for describing the long-range properties of any compressible one-dimensional system. It applies to any 1D system in which the low-lying excitations are phonons with a linear dispersion relation, $\omega(k) = ck$, where $k$ is the momentum of the excitation and $c$ is the speed of sound. 
Within LL theory, the decay of long-range correlation functions is governed by the Luttinger parameter, $K = v_F/c$, as a dimensionless ratio between the speed of sound $c$ and the Fermi velocity $v_F = \pi\hbar n/m$ where the linear density $n=N/L$ is set by the number of atoms $N$ and system size $L$. 
The many-body wave function of a Luttinger Liquid on a ring was explicitly derived within the hydrodynamic approach\cite{ReattoChester67}
\begin{eqnarray}
\psi(x_1,\cdots,x_N) = \prod_{i<j}|\sin(\pi (x_i-x_j)/L)|^{1/K}
\label{eqn:psi}
\end{eqnarray}
for large separations between particles, $|x_i-x_j|\gg \xi$, where $\xi\propto \hbar/mc$ denotes the correlation length. 
In the case of contact delta pseudopotential (Lieb-Liniger gas), the Luttinger parameter spans the range between $K\to\infty$ (ideal Bose gas) and $K=1$ (Tonks-Girardeau gas). 
For systems with finite-range interaction potentials, the values of $K<1$ become accessible, as happens for
dipoles\cite{Citro2007}, hard rods\cite{Mazzanti2008}, helium\cite{Astrakharchik14}, hydrogen\cite{Vidal16}, etc.
While the Luttinger Liquid successfully captures the long-range properties, it is inherently limited to low-energy physics and does not provide information on the full excitation spectrum.
One of few systems for which the excitation spectrum is known exactly is that of a Lieb-Liniger gas. 
Its spectrum is bounded between two branches, Lieb~I (upper, $\omega_{+}$) and Lieb~II (bottom, $\omega_{-}$) modes\cite{Lieb63}.
The Bethe {\it ansatz} technique was used to numerically evaluate its dynamic form factor~\cite{CauxCalabrese06}. 

{\it Calogero-Sutherland model}.--- The many-body wave function~(\ref{eqn:psi}) describes asymptotically 
(i.e. for $|x_{ij}|\gg\xi$) any compressible 1D system regardless of the specific interaction potential $V(x)$ between particles\cite{ReattoChester67}. 
Remarkably, there exists a special interaction potential, 
\begin{equation}
V(x)=\lambda(\lambda-1)\frac{\hbar^2}{mx^2},
\end{equation}
known as Calogero-Sutherland potential, for which wave function~(\ref{eqn:psi}) becomes the {\em exact} ground state and describes the properties of the system at {\em any} distance $x_{ij}$. Despite the existence of an exact solution, a detailed study of the dynamic form factor for the Calogero-Sutherland model is still missing, although partial results are known in the literature\cite{SimonsLeeAltshuler93,ZirnbauerHaldane95,Pustilnik_2006,ShashiPanfilCauxImambekov12}.
In the following sections, we investigate the dynamic properties of the Calogero-Sutherland model.

{\it Dynamic Structure factor}--- quantifies the linear response to the density perturbation\cite{PitaevskiiStringariBook}. At zero temperature, it characterizes the scattering cross section of inelastic reactions where the scattering probe transfers momentum $\hbar q$ and energy $\hbar\omega$ to the system \cite{LLIX}. 
Experimentally, it can be measured using Bragg spectroscopy and provides important information on the frequency response of the system.
The dynamic structure factor is directly related to the time-dependent density-density correlation function through the Fourier transformation:
\begin{eqnarray}
S(q,\omega) &=& \frac{n}{\hbar}\!\!\int\!\!\!\!\int\!\! e^{i\omega t-iqx}
\!\!\left[\frac{\langle\hat\rho(x,t)\hat\rho(0,0)\rangle}{n^2}-1\right]\!dx\;dt \notag \\
&=& \sum_\alpha \qty|\bra{\alpha}\hat\rho\ket{0}|^2\delta\qty(q-P_\alpha)\delta\qty(\omega-\omega_{\alpha 0})
\label{def Skw}
\end{eqnarray}
where a full set of many-body states $\ket{\alpha}$ with definite energy and momentum $E_\alpha, P_\alpha$ has been inserted and notation $\hat\rho =\hat \rho(0,0) $, $\hbar\omega_{\alpha 0} = E_\alpha-E_0$ is used.
The calculation of the dynamic structure factor $S(q,\omega)$ for the CSM is a long-standing challenging problem\cite{SimonsLeeAltshuler93,ZirnbauerHaldane95,Pustilnik_2006,ShashiPanfilCauxImambekov12}, due to the difficulty in evaluation of the form-factors, \emph{i.e} non-zero matrix elements $\bra{\alpha}\hat\rho\ket{0}$.

For the three special values $\lambda=1/2, 1$ and $2$, corresponding to Wigner-Dyson statistics, the structure of the excitations is as follows. 
For $\lambda=1$, excitations are the same as in an ideal Fermi gas and can be interpreted in terms of particle or hole excitations \cite{PitaevskiiStringariBook} (for spin 1/2 CSM with $\lambda=1$ see Refs~\cite{Nakai09,Nakai14}).  
For $\lambda = 1/2$, the action of the density operator generates excitations consisting of two quasi-particles and one hole. For $\lambda = 2$, it produces excitations composed of one quasi-particle and two holes. More generally, for any arbitrary rational value of the dimensionless parameter $\lambda = r/s$, the density operator acting on the ground state creates $s$ quasi-particles and $r$ holes\cite{ZirnbauerHaldane95}, which can be parametrized by their velocities (rapidities) $v_i>c$, $\bar{v}_j<c$ where $c$ is the speed of sound and  $v_{i}$, $\bar{v}_{j}$ represent the quasiparticle and quasihole velocities, respectively.
Using this parametrization of states, DSF can be expressed as a $(r+s)$-dimensional integral\cite{Pustilnik_2006}
\begin{equation}
    \label{eq:sqw_csm}
    S(q,\omega) \propto \int \prod_{i = 1}^{s} d v_{i} \prod_{j = 1}^{r} d \bar{v}_{j} F_{s , r} \delta \qty(q - P) \delta \qty(\omega - E) 
\end{equation}
where the momentum $P=P\{v,\bar{v}\}$ and energy
$E=E\{v,\bar{v}\}$ are given by Eqs.~(\ref{Eq:P}),(\ref{Eq:E}) and 
the phase space available to the elementary excitations is measured\cite{ZirnbauerHaldane95} by function $F_{s, r}$ given by\cite{Ha94}
\begin{equation}
    \label{Eq:F}
    F_{s, r} = \frac{\displaystyle\prod_{i < i'}^{s} | v_{i} - v_{i'}|^{2 \lambda} \displaystyle\prod_{j < j'}^{r} | \bar{v}_{j} - \bar{v}_{j'} |^{2 / \lambda}}{\displaystyle\prod_{\substack{i=\overline{1,s}\\j=\overline{1,r}}} (v_{i} - \bar{v}_{j} )^{2} \displaystyle\prod_{i=1}^s (v_{i}^{2} - c^{2} )^{1 - \lambda} \displaystyle\prod_{j=1}^r (c^{2} - \bar{v}^{2}_{j} )^{1 - \frac{1}{\lambda}}}.
\end{equation}
This expression was conjectured in Ref.~\cite{Haldane_conjecture} based on the results of Ref.~\cite{SimonsLeeAltshuler93} for $\lambda= 1/2$ and $2$ 
while the conjecture was proved in Refs~\cite{Ha94,Ha95} using properties of Jack polynomials.
Although conceptually straightforward, performing such an integration in practice requires specialized methods, as traditional quadrature techniques are only efficient for low-dimensional integrals. For example, discretizing each degree of freedom into 100 points for an $r+s$-dimensional integral would require evaluating $100^{r+s}$ points, which quickly becomes impractical beyond small $r$ and $s$. 
Another complication of using Eq.~(\ref{eq:sqw_csm}) is that the coefficient of proportionality is not explicitly known. In this Letter, a stochastic method is proposed for the efficient evaluation of Eq.~(\ref{eq:sqw_csm}) which automatically determines the coefficient of proportionality.

To determine the coefficient of proportionality in Eq.~(\ref{eq:sqw_csm}), we apply the $f$-sum rule\cite{PitaevskiiStringariBook} 
$\int S(q,\omega)\hbar\omega d\omega = \hbar^2q^2/(2m)$, obtaining the following expression for the dynamic form factor,
\begin{equation}
S(q, \omega ) =
\frac{
\int
\displaystyle\prod_{i = 1}^{s} d v_{i}
\displaystyle\prod_{j = 1}^{r} d \bar{v}_{j}
\left[
\frac{q^2}{2m}\right]
F_{s , r} 
\delta(q-P)\delta(\omega-E)}
{\int
\displaystyle\prod_{i = 1}^{s} d v_{i}
\displaystyle\prod_{j = 1}^{r} d \bar{v}_{j}
d\omega
\left[\omega\right] F_{s , r}\delta ( q - P) \delta ( \omega - E )}\;.
\end{equation}

To proceed with the stochastic interpretation, we represent the DSF, $S(q,\omega)$, as a ratio of two averages
\begin{eqnarray}
S(q, \omega ) = \frac{\Big\langle\delta\left(q-P\{v,\bar{v}\}\right)\delta(\hbar\omega-E\{v,\bar{v}\})\Big\rangle_{F} }{\big\langle \delta(q-P\{v,\bar{v}\}) E\{v,\bar{v}\}\Big\rangle_{F}}    
\nonumber
\end{eqnarray}

with the mean values calculated as an average over $F_{s,r}$,
\begin{equation}
\label{Eq:A}
\langle A\{v,\bar{v}\}\rangle_{F} =
\int\!\!\displaystyle\prod_{i = 1}^{s} d v_{i} \displaystyle\!\!\prod_{j = 1}^{r}\!d\bar{v}_{j}
A(v_i,\overline{v}_j) F_{s , r}  \delta ( q - P) \delta ( \omega - E ).
\end{equation}

Since $F_{s,r}$ is both real and non-negative (see Eq.~(\ref{Eq:F})), it can be treated as a probability distribution function with two imposed conditions, corresponding to having specific momentum $P$ and energy $\omega$. 
The probabilistic interpretation allows us to sample the function $F$ using the Metropolis algorithm and to evaluate the integral~(\ref{Eq:A}). 

At zero temperature $S(q,\omega)$ is different from zero for frequencies $\omega$ lying within a finite interval $\omega_{-} < \omega < \omega_{+}$ with the bound for $|q|<2k_F$ given by $\omega_{\pm} = q ( u \pm q\;/\;2m)$.
As can be seen from Eq.~(\ref{Eq:F}), the probability distribution $F_{s, r}$ exhibits a divergence for $\lambda<1$ at the edge of the quasiparticle branch where $|v_i|\to c$.
Similarly, $F_{s , r}$ diverges for $\lambda>1$ at the edge of the quasihole branch, where $|\bar v_j|\to c$. 
Such divergences pose challenges to accurate Monte Carlo sampling, which have been handled by introducing a regularized probability distribution (see End Matter). 

{\it Results.}---We numerically evaluate the form factor $S(q,\omega)$ of the CS model using the stochastic method introduced above. To our knowledge, this is the first time $S(q,\omega)$ has been determined exactly in a system exhibiting such a wide range of physical regimes, spanning from a weakly interacting Bose gas to a crystal. 
It is also important to note that for any value of $\lambda$, our results apply to both bosonic and fermionic systems. 
We present the obtained form factor as a heatmap in the $(q, \omega)$ plane in Fig.~\ref{fig1}, and we show the frequency $\omega$ dependence for characteristic values of momenta $q$ in Fig.~\ref{fig3}. 
We now discuss the features observed in different physical regimes, ranging from weak to strong interactions.

\begin{figure}[!h]
\centering
\includegraphics[width=\columnwidth]{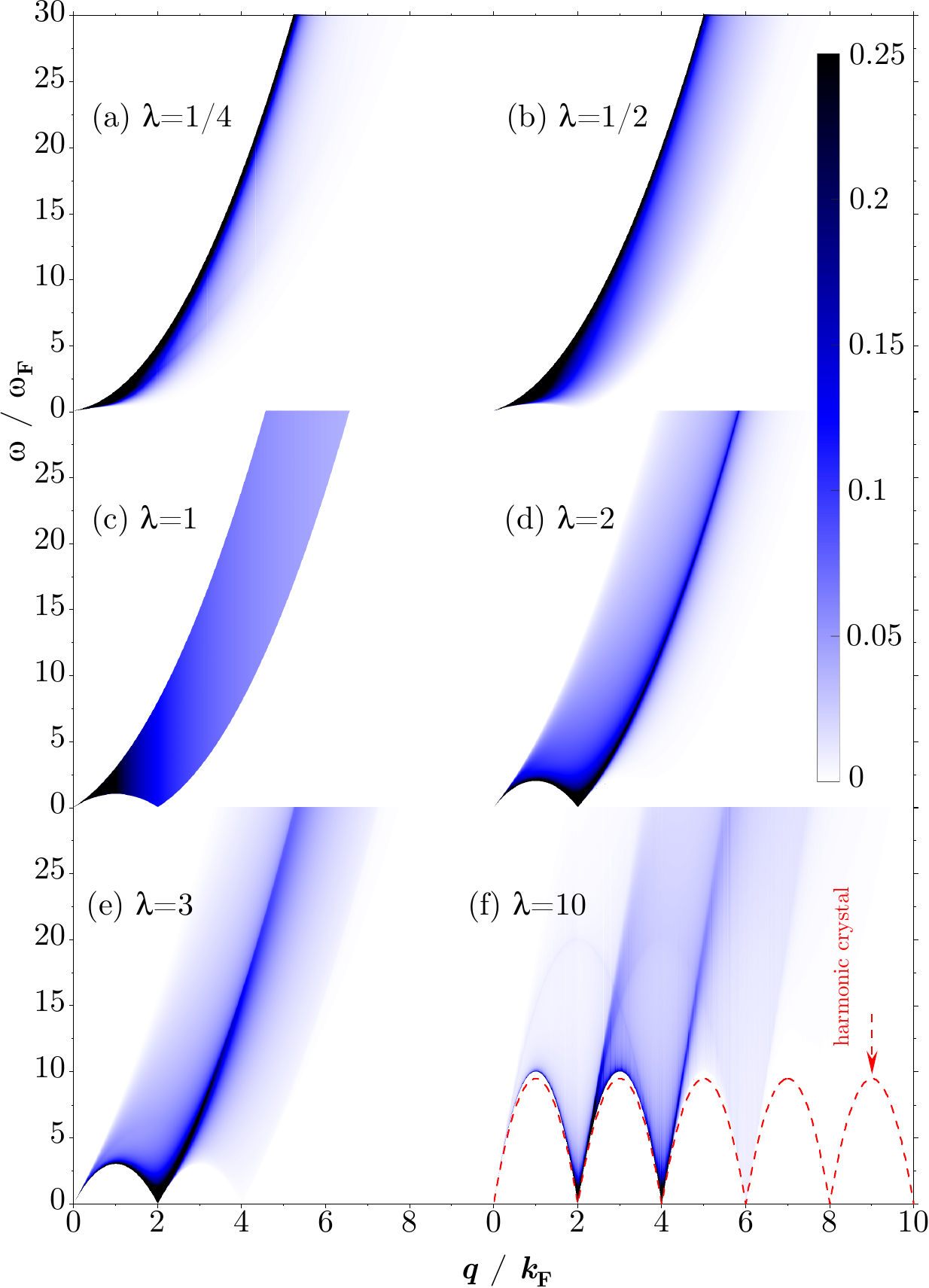}
\caption{
Dynamic structure factor $S(q,\omega)$ sampled for characteristic values of interaction parameter $\lambda$. 
The three cases (b,c,d) relevant for Wigner-Dyson distributions are shown in the central part,
(b) $\lambda=1/2$ corresponds to the critical value of the Luttinger parameter, $K_L=2$, for which a pinning transition appears;
(c) $\lambda=1$ corresponds to an ideal Fermi gas for Fermi-Dirac statistics and Tonks-Girardeau gas for Bose-Einstein statistics;
(d) $\lambda=2$ corresponds to the critical value $K_L=1/2$ at which the divergence in the momentum distribution at $k=0$ of a bosonic system disappears, so any reminiscence of a Bose-Einstein condensation is absent. 
In panel (a) we show results for $\lambda=1/4$, which corresponds to a weakly interacting system;
panels (e-f) correspond to the regime of large amplitude of the interaction potential, $\lambda$=3 and 10, and correspond to the regime in which systems form a quasicrystal.
}
\label{fig2}
\end{figure}

\begin{figure}[h!]
\centering
\includegraphics[width=\columnwidth]{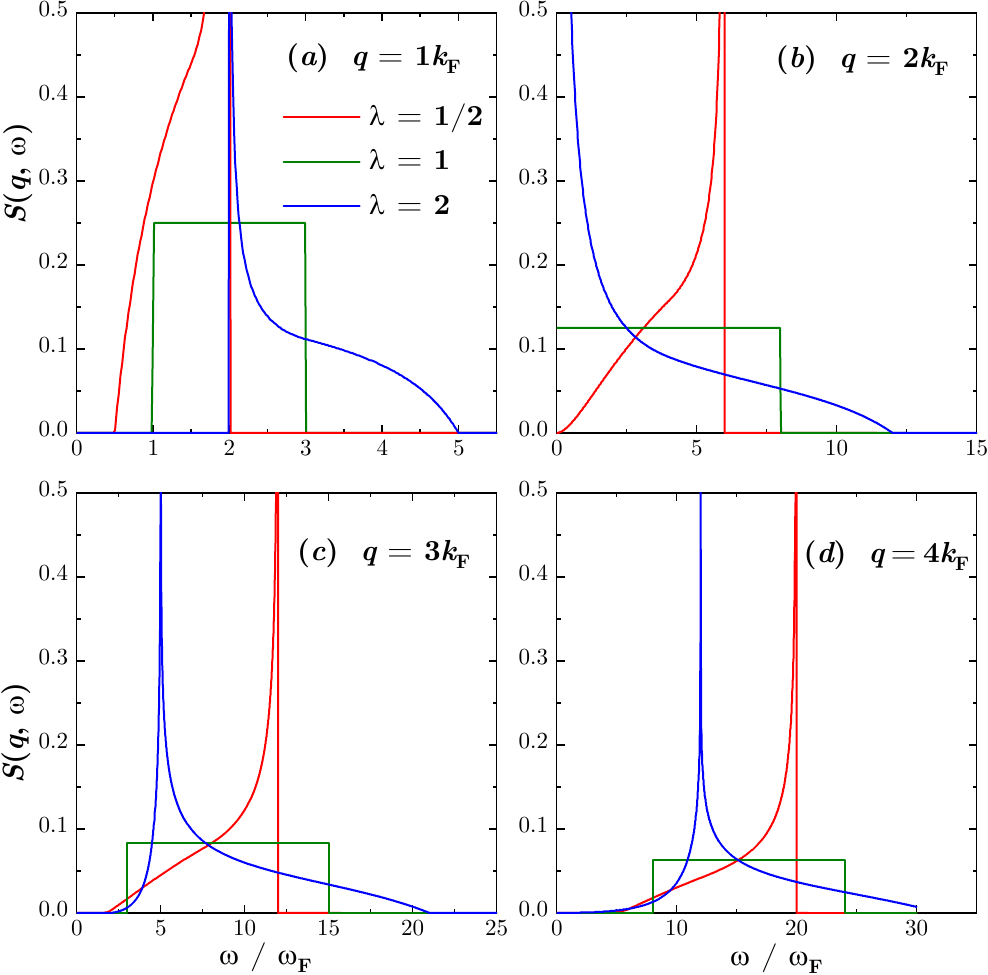}
\caption{
Frequency dependence of the dynamic structure factor $S(q,\omega)$ for four characteristic values of momenta $q/k_F = 1;2;3;4$ are shown in panels (a,b,c,d) for the three basic cases $\lambda$ = 1/2, 2, and 1, relevant to Wigner-Dyson statistics.
}
\label{fig3}
\end{figure}

For $\lambda = 0$, interactions vanish, and the considered system corresponds to either an ideal Bose gas\cite{landau1980statistical} or a Fermionic Tonks-Girardeau gas\cite{GirardeauMinguzzi06,Koscik23,Sabater24}, both having trivial local correlations, as in the bosonic case, the ground-state wave function is a constant and zero speed of sound. We do not show $S(q,\omega)$ in that trivial case. However, for any finite value of $\lambda$, there is a finite speed of sound and the low-momenta excitations correspond to phonons with linear excitation spectrum, $\omega = k c$, as can be seen in Fig.~\ref{fig2} for $k\to 0$. 

For small values of $\lambda$ (e.g., $\lambda=1/4$) shown in Fig.~\ref{fig2}, the system stays in the weakly interacting regime, and excitations are exhausted by the particle branch, $\omega_+$. There is a power-law divergence\cite{Pustilnik_2006} close to $\omega_+$, as described by Eq.~(\ref{eq:Reg}) (see End Matter). 
A similar behavior was observed\cite{CauxCalabrese06,PhysRevLett.99.110405} in a Lieb-Liniger gas in the weakly interacting regime, where $\omega_+$ coincides in this regime with the Bogoliubov dispersion\cite{Lieb63}. 
Instead, $\omega_-$ dispersion coincides with that of a gray soliton\cite{Kulish76,IshikawaTakayama80,Astrakharchik13,PustilnikMatveev14,PustilnikMatveev15} although such states are not naturally populated at equilibrium.

The case $\lambda=1/2$ is particularly interesting, as it marks the threshold at which the system begins to undergo a pinning phase transition under a commensurate periodic potential\cite{Buchler2003}. In fact, the renormalization group (RG) theory predicts\cite{GiamarchiBook} that the critical value of the Luttinger parameter is $K_L = 2$,
corresponding to $\lambda = 1/K_L = 1/2$. For bosons, this regime corresponds to strong repulsion, and for fermions to strong $p$-wave attraction. The peak is significantly broadened in the vicinity of the upper branch, while the lower branch remains essentially voided.

The simplest case in terms of excitation structure is $\lambda=1$ with $r=s=1$, where the interaction potential vanishes, $V(x)=0$. For fermions, this limit corresponds to the ideal Fermi gas\cite{PitaevskiiStringariBook}, and for bosons, to the Tonks-Girardeau gas\cite{Girardeau60}, i.e. to impenetrable bosons. The probability distribution~(\ref{Eq:F}) becomes flat, $F_{\lambda=1}=1$, indicating that all energies are equally weighted for a fixed momentum. In a stochastic formulation, this results in accepting all configurations in Monte Carlo (MC) simulations, leading to a flat $S(\omega,q)$ within the range $[\omega_{-},\omega_{+}]$ for a fixed $q$, as shown in Fig.~\ref{fig3}.

The $\lambda>1$ regime is characterized by strong correlations. 
While this regime cannot be accessed in the ground state of the Lieb-Liniger gas\cite{LiebLiniger63,CauxCalabrese06}, its initial part can be reached in the excited super Tonks-Girardeau state as predicted theoretically\cite{Astrakharchik05} and observed experimentally\cite{HallerNagerl09,KaoLev21}. 
Alternatively, $\lambda>1$ regime is naturally accessible in finite-range potentials, such as dipoles\cite{Citro2007}, helium\cite{Astrakharchik14}, hard rods\cite{Mazzanti2008,Bertaina2016}, among others. 
For fermions, it corresponds to $p$-wave repulsion\cite{Brand05}.
According to Eq.~(\ref{eq:Reg}), the divergence is shifted from the upper to the lower branch. 

Figures~\ref{fig2}-\ref{fig3} report the form factor for $\lambda=2$, which is the threshold value at which the reminiscence of Bose-Einstein condensation, seen as divergence of $k=0$ occupation, disappears in bosonic systems. 
That is, for $\lambda > 2$, the momentum distribution in a bosonic system becomes finite, and, in the ``classical'' limit $\lambda \to \infty$, the distinction between bosonic and fermionic statistics becomes increasingly negligible.
Indeed, in that limit, the system acquires many properties characteristic of classical crystals with a lattice period $a_{\text{latt}} = n^{-1}$. The very definition of a crystal, as a periodically modulated state with diagonal long-range order, requires a divergence in the static structure factor $S(q) = \int_0^\infty S(q,\omega)\,d\omega$ at $q_{\text{latt}} = 2\pi/a_{\text{latt}} = 2k_F$. 
In one dimension, the height of the Bragg peak, $S(2k_F) \propto N^{1-2K_L} = N^{1-2/\lambda}$\cite{Haldane81,KrivnovOvchinnikov82}, grows linearly with the number of particles $N$ only asymptotically, as $\lambda \to \infty$. Nevertheless, the ``quasi'' crystallization manifests as a divergence at the Bragg moment $2k_{F}$, corresponding to the {\it umklapp} processes in bosonization language\cite{Cazalilla04}. 
In the limit of large $\lambda$, the spectrum is exhausted by the lower branch $\omega_-$ and its frequency can be obtained withing harmonic crystal theory\cite{KrivnovOvchinnikov82},
\begin{eqnarray}
\label{eq:crystal}
\omega^2(q) &=& \frac{4}{m}\sum\limits_{\ell=1}^{\infty}V''\left(\frac{\ell}{n}\right)\cdot \left(1-\cos \frac{\ell q}{n}\right)\\
&=& \frac{2\lambda(\lambda-1)\hbar^2n^4}{15 m^2}
\left(\pi^4 -45\left[\text{Li}_4e^{-\frac{i q}{n}}+\text{Li}_4e^{\frac{i q}{n}}\right]\right),
\nonumber
\end{eqnarray}
where we use the polylogarithm function $L_n(z)$ to evaluate the summation. The typical Brillouin zone structure predicted by Eq.~(\ref{eq:crystal}) is asymptotically recovered for large $\lambda$ (see the characteristic case for $\lambda = 10$).

To conclude, we have introduced a stochastic method for evaluating the form factor in the Calogero-Sutherland model and have calculated $S(k,\omega)$ over a wide range of parameters, spanning from the weakly interacting regime to the ideal Fermi/Tonks-Girardeau gas, and extending into the quasi-crystal regime. 
Since the CS model can be obtained by extending the Luttinger Liquid ground-state wave function from large distances (where hydrodynamics apply) to short ones, many of the observed properties are expected to be universal. For instance, this theory provides an explanation why the lower bound $\omega_-$ should scale with $K_L$, as observed in Ref.~\cite{Bertaina2016} in helium. The advantage of the exactly solvable CSM is that it spans all possible values of the Luttinger parameter ($0<K_L<\infty$), making connections to the vast majority of one-dimensional systems that are compressible and gapless. We demonstrate that in the quasi-crystal regime, harmonic crystal theory accurately captures the lower branch $\omega_-$, which exhibits a typical Brillouin-zone structure of excitations. 
The generality of the Calogero-Sutherland model (CSM) extends its applicability beyond ultracold quantum systems. Notably, there is a connection to Random Matrix theories at three special values, $\lambda = 1/2, 1, 2$. Additionally, as pointed out in Ref.~\cite{SimonsLeeAltshuler93}, the Fourier transform of the form factor $S(r,t) = \text{FT}[S(k,\omega)]$ relates to the excitation spectra of disordered metallic grains under external perturbation, mapping position $r$ to spectral energy $\omega$ ($r \to \omega$) and time $t$ to perturbation strength $u$, with $u^2 \to 2t$, making the predictions applicable to systems in random media.

As an outlook, our approach can be used to calculate the drift force caused by dragging an impurity moved through the system\cite{Astrakharchik04,Lang15}. Furthermore, it can be extended to study a mixture of two Calogero-Sutherland (CS) gases\cite{GirardeauAstrakharchik10} or to explore the spinful Calogero-Sutherland model. In the latter case, the form factor for the simplest non-trivial scenario was computed in Ref.~\cite{Kato97}.

We acknowledge support by the Spanish Ministry of Science and Innovation (MCIN/AEI/10.13039/501100011033, grant PID2020-113565GB-C21), by the Spanish Ministry of University (grant FPU No. FPU22/03376 funded by MICIU/AEI/10.13039/501100011033), and by the Generalitat de Catalunya (grant 2021 SGR 01411). This work was supported by the following research Grant
Nos.~PID2021-127957NB-I00, PID2019-109518GB-I00, TED2021-132550B-C21 (Ministerio de Ciencia e Innovación), CEX2021-001202-M (“María de Maeztu” program for Spanish Structures of Excellence), and 2017SGR13 (Generalitat de Catalunya). 
We also acknowledge access to supercomputer resources as provided through grants from the Red Española de Supercomputación (FI-2025-1-0020).

\clearpage

\section*{\textbf{End Matter}}

\section{Mathematical Background}
The Hamiltonian of the Calogero Sutherland Model is given by\cite{sutherland2004beautiful}
\begin{equation}
\label{Eq:H}
\hat{H}
=-\frac{\hbar^2}{2m}\sum_{i=1}^{N}\frac{\partial^2}{\partial x_i^2}
+\frac{\hbar^2\lambda(\lambda-1)}{m}\sum_{i<j}^{N}
\frac{1}{\text{crd}(x_{ij})^2}
\end{equation}
where $\text{crd}(x_{ij}) = (L/\pi)\sin(\pi x_{ij}/L)$ is the chord length between particles $i$ and $j$ on a ring of circumference  $L$\cite{Lapointe1996}. 
The ground state wave function of Hamiltonian~(\ref{Eq:H}) is known exactly and can be explicitly written as
\begin{equation}
\label{Eq:wf_lambda}
\psi(x_1,...,x_N) = \prod_{i>j}|\sin \frac{\pi}{L}(x_i-x_j)|^{\lambda}
\end{equation}
with the corresponding ground-state energy equal to
\begin{equation}
\label{Eq:E_N}
\frac{E(N)}{N}=\frac{\pi^2\lambda^2}{6(1+1/N^2)}\frac{\hbar^2n^2}{m}.
\end{equation}
Equation of state~(\ref{Eq:E_N}) has a quadratic dependence on the linear density $n=N/L$, where $L$ is the length of the ring with periodic boundary conditions (PBC).

It can be shown\cite{Pustilnik_2006} that the momentum $P$ and energy $E$ of excitations relative to the ground state in Eq. (4) can be expressed in terms of the quasi-particles $v_{i}$ and quasi-holes $\hat{v}_{j}$ velocities. Some of the individual properties of the quasi-particles and quasi-holes are:
\begin{table}[h!]
\centering
\begin{tabular}{lcc}
& \qquad \underline{Quasi-particles} & \qquad \underline{Quasi-holes} \\
mass & \qquad $m$ & \qquad $\bar{m} = m/\lambda$ \\
momentum & \qquad $m v$ & \qquad $ - \bar{m} \bar{v}$ \\
energy & \qquad $\frac{m}{2}( v^{2} - c^{2} )$ & \qquad $- \frac{\bar{m}}{2} ( \bar{v}^{2} - c^{2} )$
\end{tabular}
\end{table}
\\where the speed of sound is $c = \lambda v_F$ and $v_{F}$ is the Fermi velocity defined as $v_{F} = \pi \hbar n /m$. The parameters shown above are for a rational value of $\lambda = r/s$, where $r$ is the number of \emph{quasi-holes} and $s$ number of \emph{quasi-particles}.\\
In this case, the excitations' momentum and energy are
\begin{equation}
    \label{Eq:P}
    P\{v,\bar{v}\} = \displaystyle\sum_{i=1}^{s} m ( v_{i} - c) + \displaystyle\sum_{j = 1}^{r} \bar{m} (c- \bar{v}_{j} )    
\end{equation}
\begin{equation}
\label{Eq:E}
E\{v,\bar{v}\}= c P\{v,\bar{v}\} + \sum_{i = 1}^{s} \frac{m ( v_ {i} - c )^{2}}{2} - \sum_{j = 1}^{r} \frac{ \bar{m} ( \bar{v}_{j} - c )^{2}}{2} 
\end{equation}
\\This imposes bounds on possible values of the velocities together with bounds on the frequencies of the dynamic structure factor.

\begin{table}[!h]
\centering
\begin{tabular}{cc}
\underline{Quasiparticles} & \qquad \underline{Quasiholes} \\
& \\
$c < v_{i} < v_{0}$ , $v_{0} = c + q/m$ & \qquad $\bar{v}_{0} < \bar{v}_{j} < c $ , $\bar{v}_{0} = c - q/m$ \\
& \\
Frequencies: & \qquad $\omega_{-} < \omega < \omega_{+}$ \\
Upper branch: & \qquad $\omega_{+} = c q + q^{2} / (2m)$ \\
Lower branch: & \qquad $\omega_{-} = c q - q^{2} / (2\bar{m})$ 
\end{tabular}
\end{table}
\vfill\null
The upper branch describes particle excitation and has no limitations on the transferred moment $q$. The lower branch corresponds to promoting a particle to the Fermi surface and effectively is a hole excitation. For a single-hole excitation, the transferred momentum is limited to $|q|<2k_F$.
For momenta larger than that, the lowest energy is generated by several {\it umklapp} excitations and a hole excitation. Each {\it umklapp} excitation changes the momentum by $2k_F$ with no additional cost to the energy. In this way, the lower bound within $n$-th Brillouin zone is $\omega_{-} = c \mod(q,2k_F) - \mod(q,2k_F)^{2} / \bar{m}$, where $\mod(q,2k_F) = q - n2k_F$.

\section{Monte-Carlo Algorithm}
Following a Monte-Carlo sampling, we obtain $S (q, \omega)$ for a given momentum $q$ and a certain value of interaction parameter $\lambda$ by calculating an integral of $r + s$ dimensions. We interpret $F_{s, r} > 0$ as a probability distribution and sample it using the MC method \cite{Landau_2005}.
\\While divergences are expected in Eq.(7) for some critical $\lambda$ values, discretized calculations' inherent nature does not allow rigorously capturing them, resulting in finite values where infinity divergences are expected. Therefore, to address this issue, we employ a reweighting technique where, instead of sampling divergent $F_{s,r}$ directly, we sample the regularized probability distribution $F_{s,r}/\text{Reg}(\omega,\lambda)$ and average $\text{Reg}(\omega,\lambda)A(v_{i},\bar{v}_{j})$ rather than $A(v_{i},\bar{v}_{j})$. We chose the regularizing function as 

Therefore, to address this issue, we employ the reweighting technique applied to Eq.~(\ref{Eq:A}).
Instead of sampling divergent $F_{s , r}$ directly, we sample the regularized probability distribution $F_{s , r} / Reg(\omega,\lambda)$ and average $Reg(\omega,\lambda)A(v_i,\overline{v}_j)$ rather than $A(v_i,\overline{v}_j)$. We chose the regularizing function as
\begin{equation}
    \label{eq:Reg}
\text{Reg}(\omega,\lambda) =
\begin{cases}
|\omega- \omega_{-}|^{(\frac{1}{\lambda}-1)} & \quad \text{if} \quad \lambda > 1\\
|\omega_{+} - \omega|^{(\lambda -1)} & \quad \text{if} \quad \lambda \leq 1   
\end{cases}
\end{equation}
so that the divergences are removed from the probability distribution function, ensuring more accurate MC sampling.

Our Monte Carlo algorithm consisted of generating the histogram $H(\omega)$, for a fixed moment $q$, within the frequency range defined in the previous section. For each $q$ value, we sampled $N_{times}$ times our MC samplings with $N_{MC}$ steps on every time. For the histogram, we employed $N_{bins}$ for defining the number of $\omega$ subdivisions in our histogram within the range $\omega_{-} < \omega < \omega_{+}$. For a fixed moment $q$, we normalized the obtained histogram after all $N_{times} \dot N_{MC}$ samplings as $H'_{i} = \frac{H_{i}}{\sum_{i} H_{i}}$ and compute the dynamic structure factor as:
\begin{equation}
    \label{Eq:S_Reg}
    S(\omega,q) = \text{Reg}(\omega,\lambda) \frac{1}{2}\frac{q^{2} H[\omega]}{\delta \omega \bar{\omega}}
\end{equation}
    
were $\delta \omega$ is the histogram binning and $\bar{\omega}$ is the mean frequency computed as $\bar{\omega} = \sum_{i} \omega_{i} \text{Reg}(\omega_{i},\lambda)$.

When sampling different values of $\omega$, to keep the momentum constant while updating the velocities, the changes are introduced by pairs with the following two possibilities (here $\delta$ is a random number):  
\begin{itemize}
    \item $v'_{i} = v_{i} + \delta$ $\rightarrow$ $v'_{j} = v_{j} - \delta$ if the two velocities correspond to two quasi-particles or two quasi-holes.
    \item $v'_{i} = v_{i} + \delta$ $\rightarrow$ $\bar{v}'_{j} = \bar{v}_{j} + \lambda \delta$ if the two velocities are from different kinds.
\end{itemize}

The results from Fig.2 have been obtained using $N_{MC} = 10^{7}$ and $N_{blocks} = 300$. We used the same grid for moment and histogram, $N_{hist} = 1000$, and the corresponding ranges were $q \in ( 0.01,10]$ and $\omega \in [0,\omega_{max}]$. The optimal \emph{sampling amplitude} $a$ have been obtained by sampling different $a$ values for fixed parameters using $\lambda = 1$, for which all energies are expected to have equal acceptance probability within the energy range until a flat histogram has been obtained for $N_{hist} = 1000$. The resultant optimal value is $a = 1$. However, a probability-dependent adaptive method has been implemented to ensure a minimum acceptance rate of 0.5 in more complex systems. 
For Fig.3, the used parameters were $N_{MC} = 5 \cdot 10^{7}$, $N_{blocks} = 2500$ for $\lambda = \{ 1,2, \text{and } 1/2 \}$; and $N_{blocks} = 2800$ for $\lambda = \{ 3,10, \text{and } 1/4 \}$. The resolution of the energy histogram has been increased to $N_{hist} = 2000$.

\end{document}